\documentclass[aps,pra,groupedaddress,superscriptaddress,nofootinbib]{revtex4}

\usepackage{graphicx,amsmath,amssymb,amsbsy,subfigure,hyperref,bbm,times,txfonts,float}
\usepackage{subfigure,hyperref,bbm,times}
\usepackage[T1]{fontenc}
\usepackage{braket}
\usepackage{epsfig} 
\usepackage{color}
\usepackage{graphicx}
\usepackage{dcolumn}
\usepackage{bm}

\providecommand{\openone}{\leavevmode\hbox{\small1\kern-3.8pt\normalsize1}}

\usepackage{soul}

\hypersetup{
   colorlinks=true,
   linkcolor=blue,
}
\graphicspath{{figure/}}

\begin{document}

\title{Enhancing Quantum Synchronization in a driven qubit system coupled to a structured environment}
\author{Amir Hossein Houshmand Almani}
\affiliation{Department of Physics, University of Guilan, P. O. Box 41335-1914, Rasht, Iran}
\author{Ali Mortezapour}
\email{mortezapour@guilan.ac.ir}
\affiliation{Department of Physics, University of Guilan, P. O. Box 41335-1914, Rasht, Iran}
\author{Alireza Nourmandipour}
\affiliation{Department of Physics, Sirjan University of Technology, 7813733385 Sirjan, Iran}

\begin{abstract}

In this paper, we delve into the issue of Quantum Synchronization in a driven two-level (qubit) system situated within a structured environment. Our findings have practical implications as we discover that adding a time-dependent periodic modulation to the transition frequency of the qubit can significantly enhance quantum synchronization. We first discovered the phase preference and, consequently, the phase locking conditions in our system using the Husimi $Q$-function. It is revealed that combining frequency modulation and non-Markovian effects enables us to achieve a stable phase-locking for the system. We show that tuning the amplitude-to-frequency ratio of the modulation process on the zeros of the zeroth-order Bessel function led to phase locking and, thus, surprisingly enhances quantum synchronization in the system. These results provide new insights into efficiently understanding phase dynamics in quantum environments.
\end{abstract}

\date{\today }

\maketitle

\section{INTRODUCTION}
Synchronization is a phenomenon in which multiple bodies adjust their movement and rhythm through mutual interaction. This phenomenon is a prevalent behavior in nature, significant in physics, chemistry, biology, medicine, and technology. For instance, the Belousov-Zhabotinsky reaction known as Oscillating chemical reactions serves as a basis for synchronization studies to simulate the behavior of biological organisms \cite{Taylor2009-jn}. Pacemaker cells synchronize to produce our heartbeat \cite{Michaels1987-tv}, and in general, synchronization plays an influential role in various physiological rhythms \cite{Glass2001-si}. Moreover, the technological applications of synchronization involve improving the frequency stability of lasers \cite{1077265}. The instances mentioned thus far are grounded on the principles of classical physics. Nevertheless, the idea has recently surfaced that synchronization can occur in quantum systems. While quantum synchronization is compatible with classical physics, it is not as straightforward as its counterpart. In this way, quantum synchronization is analogous to classical synchronization but governed by quantum mechanical principles. Such a concept arises when two or more quantum systems adjust their dynamics to achieve correlated behavior, despite the uncertainties and the probabilistic nature of quantum mechanics.
\\
The study of quantum synchronization not only deepens our understanding of collective quantum behavior but also holds promise for applications in quantum metrology \cite{Vaidya2024-hi}, quantum communication \cite{Azahari2024-zn}, and the development of quantum networks \cite{Zhang2023-bd,Spiess2023-ez}. There is evidence that synchronized quantum systems could make quantum sensing devices more accurate at measuring things \cite{Vaidya2024-hi} or make it easier to build large-scale quantum information processing networks \cite{Zhang2023-bd}.
\\
In recent years, the study of quantum synchronization in open quantum systems has attracted the attention of some researchers \cite{PhysRevA.95.043807,PhysRevA.101.062104,PhysRevA.103.062217,AnnPhys.533.2100038,arXiv.2205.08822,PhysRevA.107.022221,arXiv.2311.05664,PhysRevA.105.062206,CommTheorPhys.73.105101,SciRep.9.19933}. Such efforts could address fundamental questions about how environmental interactions can facilitate or hinder synchronization processes at the quantum level. It is demonstrated that a two-level system is capable of exhibiting all the signatures of synchronization, stemming from the limit cycles present in the unperturbed dynamics \cite{PhysRevA.101.062104}. Karpat et al. explored the relationship between the degree of non-Markovianity and the spontaneous synchronization of a pair of qubits \cite{PhysRevA.103.062217}. In another work \cite{arXiv.2205.08822}, the authors discussed the synchronization of a single qubit interacting with a non-Markovian environment, highlighting that phase locking can occur outside the Arnold tongue region. In addition, the effect of a classical driving field on the synchronization dynamics of a qubit system within a Lorentzian \cite{PhysRevA.107.022221}, and Ohmic reservoir \cite{arXiv.2311.05664} is investigated.
\\
Inspired by these studies, in this work, we aim to examine how frequency modulation can affect the synchronization dynamics of a qubit system embedded in a leaky cavity. Generally, a quantum system experiences frequency modulation when its energy levels are changed by external driving. Frequency modulation in an atomic qubit can be performed by applying an external off-resonant field \cite{PhysRevA.58.2265,PhysRevA.68.025401,RepProgPhys.80.056002}. On the other hand, frequency modulation is now possible in superconducting Josephson qubits (artificial atoms), which are the preferred building blocks of contemporary quantum computer prototypes \cite{Nature.543.S1}, thanks to recent experimental advancements in the fabrication and control of quantum circuit-QED devices \cite{RepProgPhys.80.056002,PhysRevLett.87.246601,Science.310.1653,PhysRevLett.105.257003,NatCommun.4.1420}. It was shown that frequency modulation of a qubit can induce sideband transitions \cite{PhysRevA.86.022305,PhysRevB.87.220505}, modify its fluorescence spectrum \cite{PhysRevA.64.013813}, alter population dynamics \cite{JPhysB.41.065501,PhysRevA.61.025802,PhysRevA.80.062109,PhysRevLett.115.133601,PhysRevA.90.043838}, amplify non-Markovianity \cite{EPL.118.20005,SciRep.8.14304}, and prolong quantum resources \cite{SciRep.8.14304,QuantInfProc.22.254}. Furthermore, external control of frequency of the qubit has facilitated the identification of an accurate relationship between non-Markovianity and quantum speed limit time (QSLT) \cite{PhysRevA.106.062431}.
\\
The paper is structured as follows: In Sec. \ref{model}, we present the model and discuss the state evolution of the system. In Sec. \ref{sec.RD}, we investigate the features of synchronization of a frequency modulated qubit by calculating the Husimi $Q$-function. We then introduce a synchronization measure and the discuss how the synchronization region is affected by frequency modulation factors. finally, the main conclusions are represented in Sec. \ref{sec.iv}.

\section{The Model}
\label{model}
Our proposed model consists of a single qubit enclosed within a high-Q cavity. This cavity contains quantized modes and operates in a zero-temperature environment. In Figure 1, it is postulated that the qubit with transition frequency $\omega_0$ containing ground and excited states $\ket{g}$ and $\ket{e}$ , respectively, exhibits distinct transition frequencies that are varied sinusoidally by an external driving field, a crucial component in the operation of the model. The total Hamiltonian of the system in the dipole and rotating wave approximations is expressed as ($\hbar=1$):
\\
\begin{equation}
\label{Eq1}
\hat{H}=\hat{H}_{\text{S}}+\hat{H}_{\text{R}}+\hat{H}_{\text{int}},
\end{equation}
in which,
\begin{equation}
\begin{aligned}
\hat{H}_{\text{S}}&=\frac{1}{2}[\omega_0+d\cos(\Omega t)]\hat{\sigma}_z\\
\hat{H}_{\text{R}}&=\sum_{k}\omega_{k} {\hat{a}_{k}}^{\dagger}\hat{a}_{k}\\
 \hat{H}_{\text{int}}&=\sum_{k}\left( g_{k}\hat{\sigma}_+\hat{a}_{k}+g_{k}^*\hat{\sigma}_-\hat{a}_{k}^\dagger\right) , 
\end{aligned}
\end{equation}
where $d$ and $\Omega$ denoting the modulation amplitude and frequency, respectively, and ${{\hat{\sigma}}_{z}}=\left| e \right\rangle \left\langle  e \right|-\left| g \right\rangle \left\langle  g \right|$ is the population inversion operator of the qubit.
${{\omega }_{k}}$ represents the frequency of  the cavity quantized modes while ${{\hat{a}}_{k}}$ ($\hat{a}_{k}^{\dagger}$) are the annihilation (creation) operators of the cavity $k$th mode.The raising (lowering) operator for the qubit is denoted by ${{\hat{\sigma }}_{+}}=\left| e \right\rangle\left\langle  g \right|$ (${{\hat{\sigma }}_{-}}=\left| g \right\rangle\left\langle  e \right|$). Moreover, ${{g}_{k}}$ represents the coupling strength between  the cavity modes and the qubit.

 \begin{figure}[ht]
	\centering
	\includegraphics[width=0.4\textwidth]{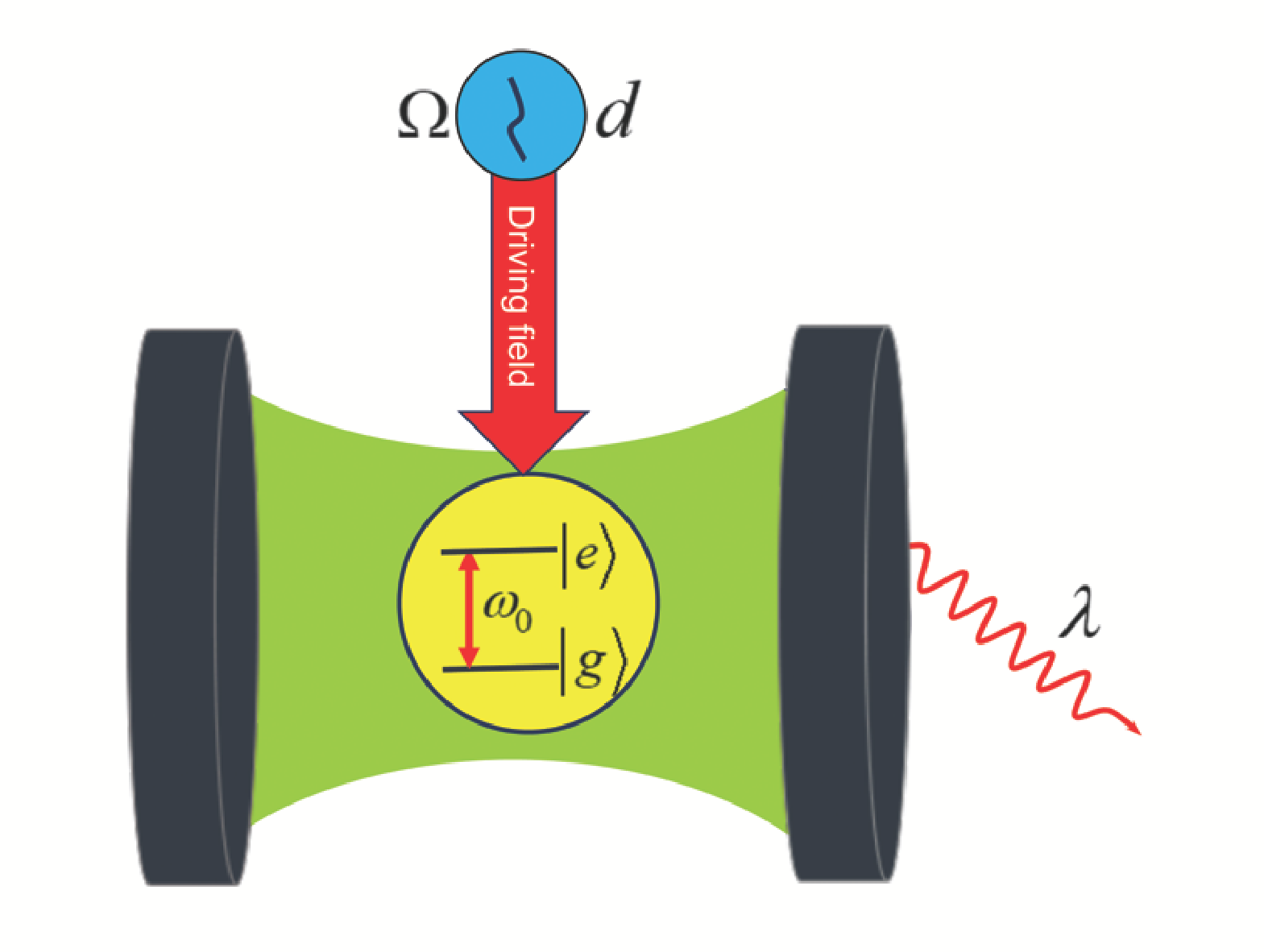}
	\caption{\label{Fig1} A schematic illustration of a qubit with a transition frequency $ \omega_{0} $ is embedded in a high-Q leaky cavity. The photon losses of the cavity, characterized by the spectral width $ \lambda \sim \frac{1}{Q} $ of the coupling, are assumed to follow a Lorentzian spectrum. A notable feature of this setup is the sinusoidal modulation of the transition frequency of the qubit by an externally applied field with a modulation amplitude $ d $ and a modulation frequency $ \Omega $. The qubit interacts with vacuum modes}
\end{figure}

According to Ref. \cite{FEI200577}, entanglement (in fact, any quantifier of it) remains invariant under local unitary transformations of the form  ${\cal U}_1\otimes{\cal U}_2\otimes\cdots\otimes{\cal U}_n$ \cite{FEI200577}. Therefore, we consider the local unitary transformation of the form
\begin{equation}
\begin{aligned}
\hat{U}&=\overrightarrow{T}\exp{\left[-i\int_0^t\left( \hat{H}_{\text{S}}(\tau)+\hat{H}_{\text{R}}(\tau)\right) d\tau\right] }\\ 
& =\exp{}\left[ -i\left\lbrace \frac{1}{2}[\omega_0t+(d/\Omega)\sin(\Omega t)]\hat{\sigma}_z + \sum_{k}\omega_{k} {\hat{a}_{k}}^{\dagger}\hat{a}_{k}t\right\rbrace \right] 
\end{aligned}
\end{equation}
where, $\overrightarrow{T}$ is a time-ordering operator. Under which, the effective Hamiltonian $\hat{H}_{\text{eff}}=\hat{U}^{\dagger}\hat{H}\hat{U}+i(\partial\hat{U}^{\dagger}/\partial t)\hat{U}$ is obtained:
\begin{equation}
\label{HEff}
\begin{aligned}
\hat{H}_{\text{eff}}&=\sum_{k}\left( \hat{\sigma}_{+}g_{k}\hat{a}_{k}e^{-i(\omega_k-\omega_0)t}e^{i(d/\Omega)\sin\Omega t}
+\text{H.c.}\right).
\end{aligned}
\end{equation}
Making use of Jacobi-Anger expansion, the exponential factors in Eq. (\ref{HEff}) can
be written as
\begin{equation}
\label{Eq5}	
e^{\pm i(d/\Omega)\sin\Omega t}=J_0\left( \frac{d}{\Omega}\right)+2\sum_{n=1}^{\infty}(\pm i)^nJ_n\left( \frac{d}{\Omega}\right)\cos(n\Omega t),
\end{equation}
where, $J_n\left( \frac{d}{\Omega}\right)$ is the $n$-th Bessel function of the first kind.

With the initial system as $\ket{\psi_0}=(C_0\ket{g}+C_1\ket{e})\otimes\ket{0}_R$, the state at any time $t$ is
\begin{equation}
\label{Eq.Psit}
\ket{\psi(t)}=C_0\ket{g}\otimes\ket{0}_R+C_1B(t)\ket{e}\otimes\ket{0}_R+\sum_kB_k(t)\ket{g}\otimes\ket{1_k}_R
\end{equation}
where, $\ket{0}_k$ is the vacuum state and $\ket{1_k}$ is the state with only one photon in the mode $k$ of the environment. Then, by substituting $\ket{\psi(t)}$ into the time-dependent Schr\"{o}dinger equation, one arrives at the following integro-differential equation for the amplitude $B(t)$:
\begin{equation}
\label{Eq.Bdot}
\dot{B}(t)+\int_0^tK(t,t')B(t')dt'=0
\end{equation}  
in which the kernel $K(t,t')$ in the continuous limit of the environment frequency takes the following form:
\begin{equation}
K(t,t')= \frac{\gamma\lambda}{2}e^{-\lambda(t-t')}\exp\left[ i\left( \frac{d}{\Omega}\right) \left\lbrace \sin\Omega t-\sin\Omega t'\right\rbrace \right] 
\label{Kernel}
\end{equation}

The explicit form of the qubit reduced density matrix at any time $t$ is derived by tracing over the environmental variables. In the computational basis, it takes the following form:
\begin{equation}
\hat{\rho}(t)=\begin{pmatrix} \rho_{ee}(0)|B(t)|^2 && \rho_{eg}(0)B(t)\\
\rho_{ge}(0)B^*(t) && 1-\rho_{ee}(0)|B(t)|^2
\end{pmatrix},
\label{QDenMat}
\end{equation}
where $\rho_{ee}(0)=|C_1|^2$, $\rho_{eg}(0)=C_1C_2^*$ and $\rho_{ge}(0)=C_1^*C_2$.

\section{Results and discussions}
\label{sec.RD}

\subsection{Husimi function and synchronization}

Our investigation of synchronization phenomena employs the Husimi $Q$-function formalism as a powerful tool for quantifying and visualizing the system's phase preferences. The Husimi $Q$-function, a quasi-probability distribution function, provides an elegant framework for representing the phase space of quantum systems \cite{scully1997quantum,PhysRevA.107.022221}. For a qubit system, the Husimi $Q$-function is formally defined as:
\\

\begin{equation}
Q(\theta,\phi,t)=\frac{1}{2\pi}\bra{\theta,\phi}\rho(t)\ket{\theta,\phi}
\label{QFunction}
\end{equation}

where, $\ket{\theta,\phi}=\cos(\theta/2)\ket{e}+\sin(\theta/2)\exp(i\phi)\ket{g}$ are the eigenstates of the operator  ${\hat{\sigma}} \cdot {n}$ with ${n} = (\sin\theta \cos\phi, \sin\theta \sin\phi, \cos\theta)$.It is straightforward to reach the explicit form of the Husimi $Q$-function as follows

\begin{equation}
Q(\theta,\phi,t)=\frac{1}{2\pi}\left(\cos(\theta) \rho_{ee}(0)|B(t)|^2 + \sin(\theta) \Re\left( e^{i\phi}\rho_{eg}(0)B(t)\right) + \sin^2(\theta/2)\right). 
\end{equation}
\\

\begin{figure}[h!]
	\centering
	\includegraphics[width=0.7\textwidth]{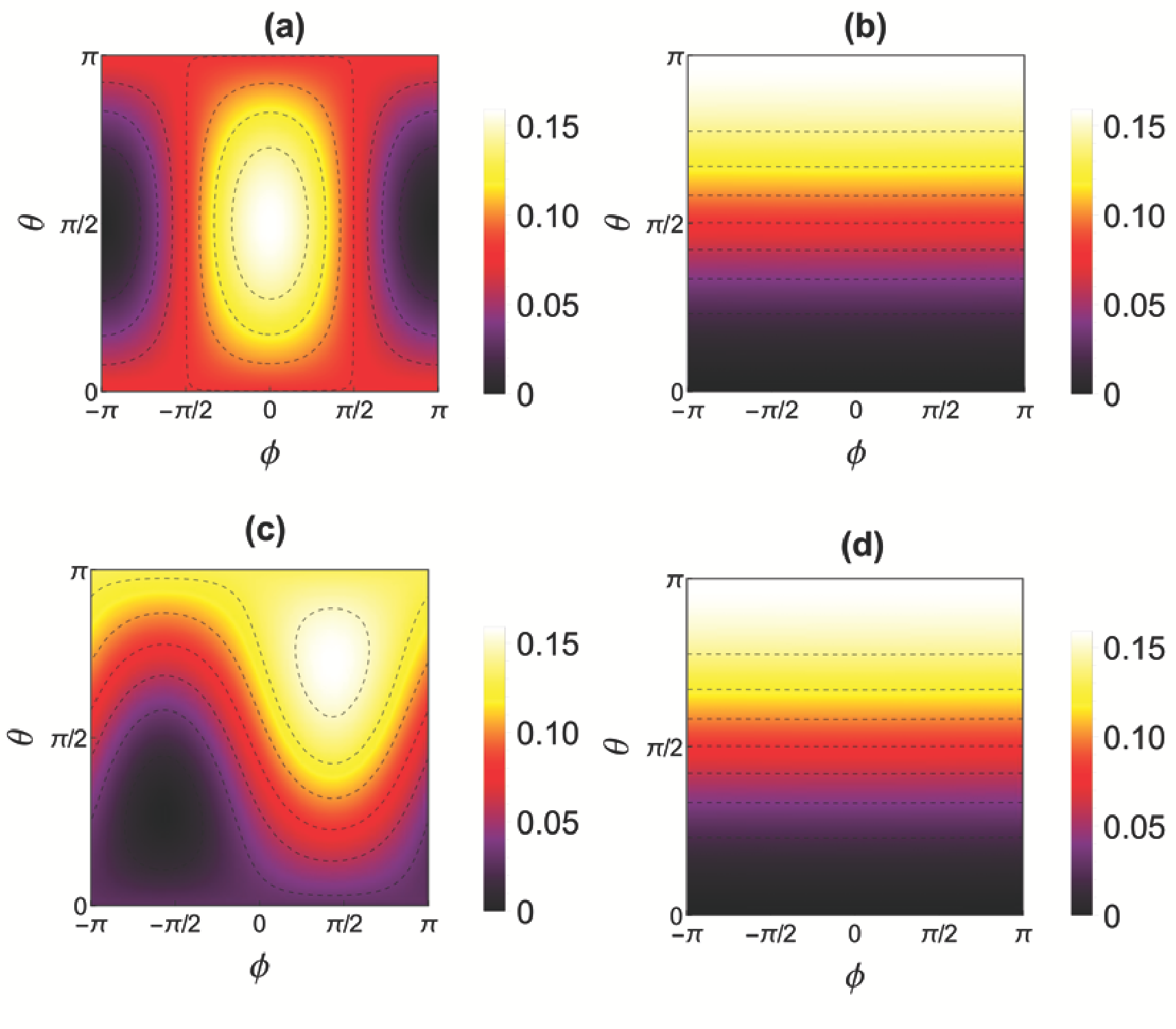}	
	\caption{The Husimi $Q$-function as functions $\theta$ and $\phi$ for (a) $\Omega = 0$, $\gamma t = 0$, (b) $\Omega = 0$, $\gamma t = 10$, (c) $\Omega = 0.001\gamma$, $\gamma t = 10$, (d) $\Omega = 100\gamma$, $\gamma t = 10$. Other parameters are $\lambda=3\gamma$ (the weak coupling regime) and $d=10\gamma$.} \label{Fig2}
\end{figure}

We consider a comparative analysis of the system dynamics in both strong and weak coupling regimes, with particular emphasis on the phase locking phenomenon.Figure \ref{Fig2} displays the $Q$-function as functions of $\theta$ and $\phi$ for different values of the scaled time $\gamma t$ and the frequency modulation $\Omega$ in the weak coupling regime. For the initial state of the system at time $\gamma t$=0, the quasi-probability distribution has its maximum value at $\phi$=0, indicating a non-uniform phase distribution. This non-uniformity expresses that the two-level system at the initial time $\gamma t$=0 has an initial phase preference at $\phi$=0. It is observed that with time and in the absence of the modulation process, the initial phase preference in the $Q$ distribution function fades and eventually reaches a steady state without any oscillations. By performing the modulation process, oscillations appear in the $Q$-function such that these oscillations are more significant for more minor frequencies. As the modulation frequency increases, the intensity of these oscillations decreases and reaches the state that previously existed without modulation frequency. The smaller the modulation frequency, the larger the oscillations are, and the longer it takes for the $Q$-function to reach the steady state. However, when we increase the modulation frequency, the system reaches the steady state faster, and the oscillations are also reduced. The reason for such oscillations is the induction of non-Markovian features in the weak coupling regime by the frequency modulation process. As shown in Figure \ref{Fig2}(d), for large modulation frequencies ($\Omega$ = $100\gamma$), the $Q$-function reaches a steady state similar to the unmodulated situation. The key conclusion drawn from this is the absence of phase locking in the weak coupling regime.
\\

\begin{figure}[h!]
	\centering
	\includegraphics[width=0.9\textwidth]{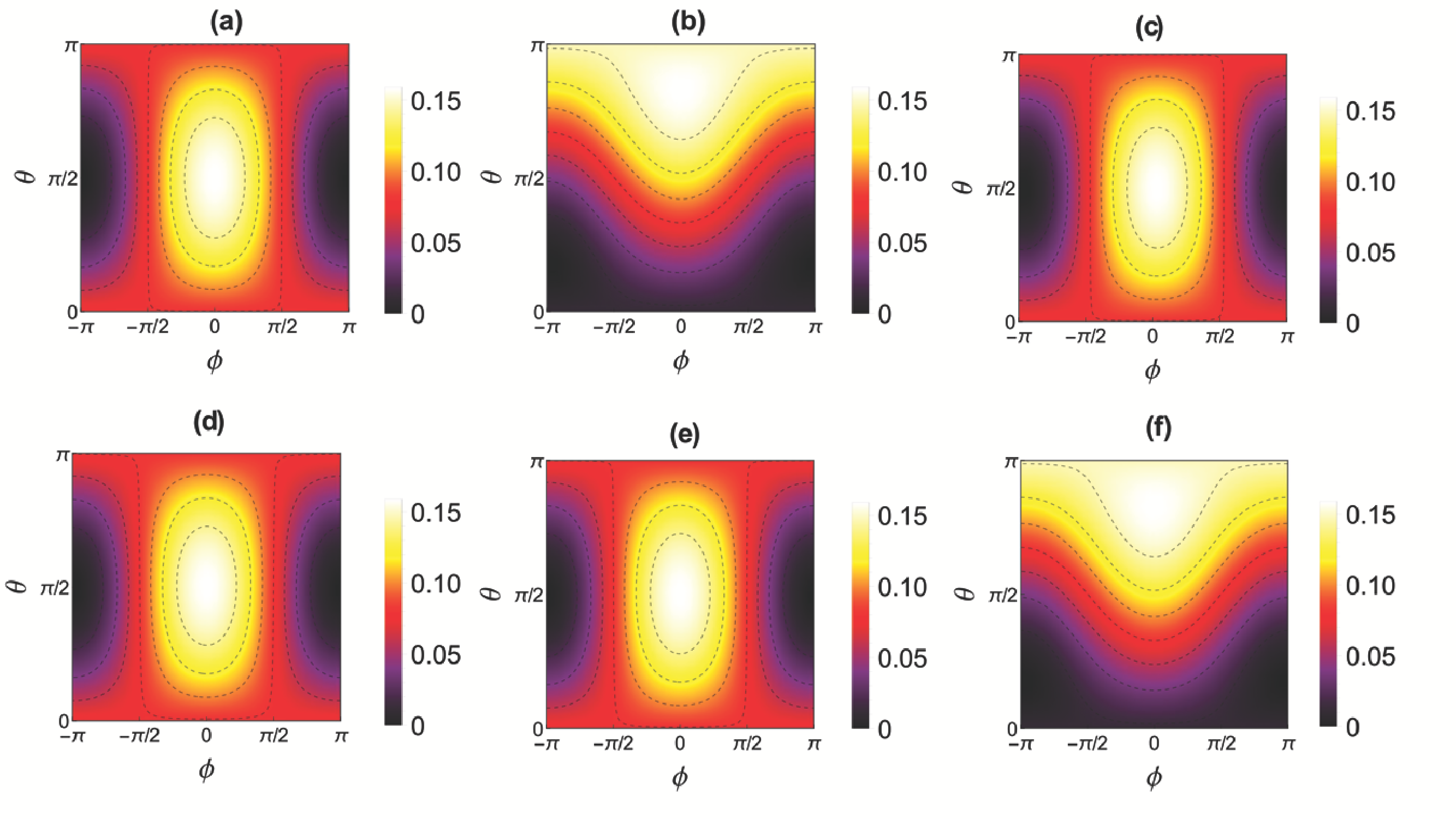}	
	\caption{The Husimi $Q$-function as functions $\theta$ and $\phi$ for (a) $\Omega = 0$, $\gamma t = 0$, (b) $\Omega = 0$, $\gamma t = 100$, (c) $\Omega = 0.001\gamma$, $\gamma t = 100$, (d) $\Omega = 0.9\gamma$, $\gamma t = 100$, (e) $\Omega = 2.1\Omega$, $\gamma t = 100$, (f) $\Omega = 50\Omega$, $\gamma t = 100$. Other parameters are$\lambda=0.01\gamma$ (the strong coupling regime) and $d=5\gamma$.} \label{Fig3}
\end{figure}

Figure \ref{Fig3} exhibits the $Q$-function as functions of $\theta$ and $\phi$ for different values of the scaled time $\gamma t$ and the frequency modulation $\Omega$ in the strong coupling regime. Without a modulation process, phase information is lost over time, a phenomenon that takes longer compared to a similar case in the weak coupling regime. By comparing Figures \ref{Fig2}(b) and \ref{Fig3}(b), it is deduced that in the weak coupling regime, the $Q$-function directly reaches the steady state. while, in the strong coupling regime, it is accompanied by oscillations. On the other hand, it is observed that for large modulation frequencies $\Omega  =50\gamma$), the $Q$-function reaches the steady state similar to the situation where frequency modulation does not occur (see Figure \ref{Fig3}(f)). In Figure \ref{Fig3}(c), when the modulation frequency is small($\Omega$ = $0.001\gamma$), it is seen that phase locking occurs in the range $\gamma t<300$, and as a result, the initial phase preference is preserved. However, the interesting results occur for $\Omega$ = $0.9\gamma$ and $\Omega$ = $2.1\gamma$. Here, we observe that the initial phase preference in the $Q$-function is preserved for a more extended period of time, leading to phase locking.
\\

\begin{figure}[h!]
	\centering
\includegraphics[width=0.9\textwidth]{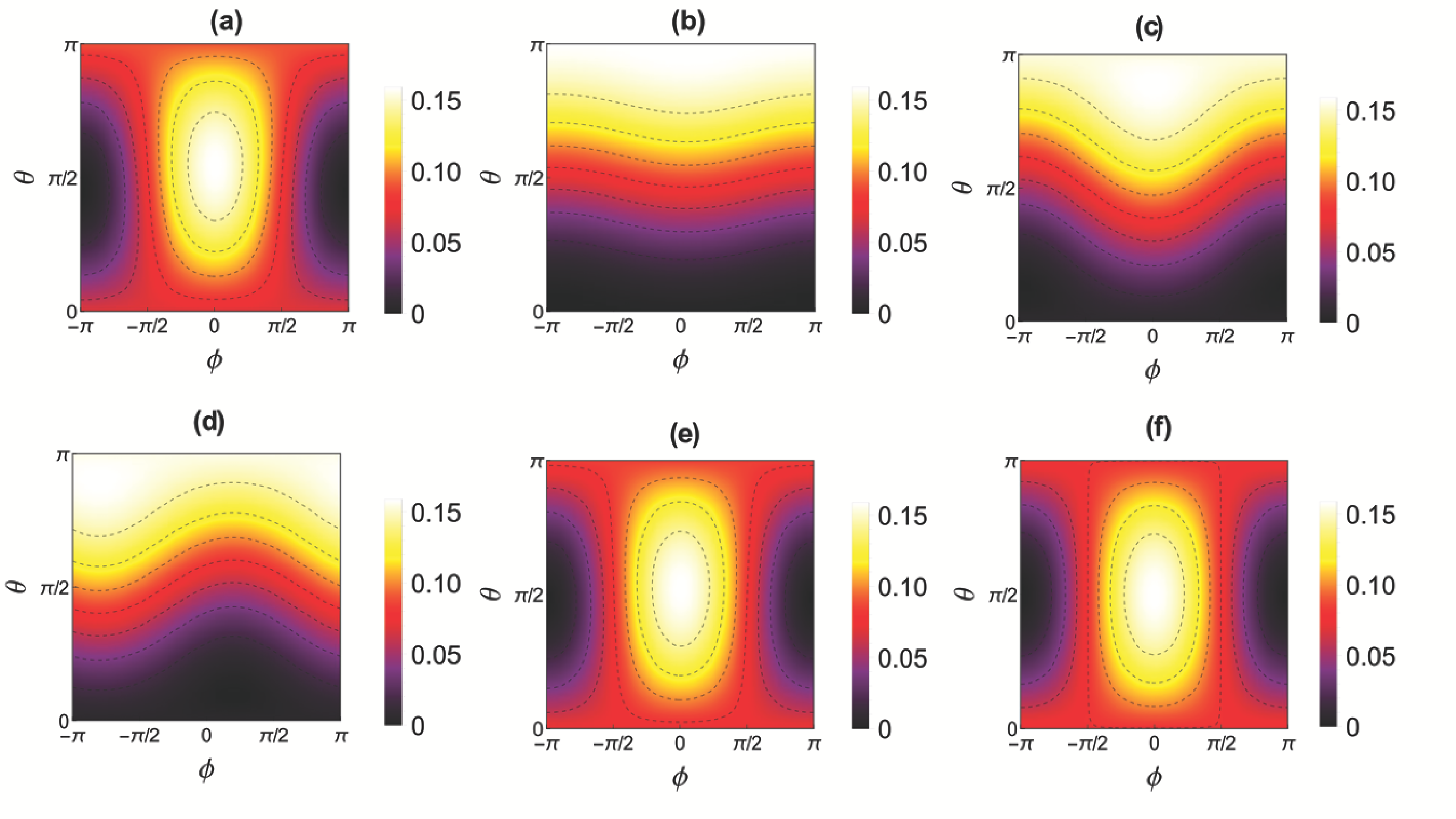}
	\caption{The Husimi $Q$-function as functions $\theta$ and $\phi$ for (a) $\Omega = 0.05\gamma$, $\gamma t = 100$, $d = 5\gamma$, (b) $\Omega = 0.5\gamma$, $\gamma t = 100$, $d = 5\gamma$, (c) $\Omega = 5\gamma$, $\gamma t = 100$, $d = 5\gamma$, (d) $\Omega = 0.05\gamma$, $\gamma t = 100$, $d = 2.40483\Omega$ [$J_0(2.40483) = 0$], (e) $\Omega = 0.5\Omega$, $\gamma t = 100$, $d = 2.40483\Omega$ [$J_0(2.40483) = 0$], (f) $\Omega = 5\gamma$, $\gamma t = 100$, $d = 2.40483\Omega$ [$J_0(2.40483) = 0$] in the strong coupling regime with $\lambda=0.01\gamma$ } \label{Fig4}
\end{figure}

To gain a more comprehensive understanding of the system evolution, we now meticulously study the interplay of the two driving parameters $d$ and $\Omega$, which is expected to affect phase locking significantly. In this regard, in Figure \ref{Fig4}, the $Q$-function is drawn in terms of $\theta$ and $\phi$ for different values of the modulation frequency under a strong coupling at $\gamma t = 100$. In the first row (Figure \ref{Fig4}(a-c)), it is assumed $d=5\gamma$, and in the second row (Figure \ref{Fig4}(d-f)), we have set the ratio $d$/$\Omega$ at the first zero of the Bessel function $J_0$ ($d$/$\Omega$= 2.40483). Comparing these two rows reveals that by increasing the $\Omega$ and adjusting the ratio $d$/$\Omega$ on the first zero of the Bessel function ($J_0$), the preservation of the initial phase preference in the $Q$ distribution function is significantly improved. This improvement in maintaining the initial phase preference is crucial in enhancing the quality of phase locking. The main reason for the success of this method lies in the fact that when the $d$/$\Omega$ ratio is adjusted so that $J_0$ becomes zero, increasing the modulation frequency creates a robust protective mechanism against noise. This protective mechanism significantly improves the system’s stability against external disturbances.
\\

\subsection{{Synchronization Measure $S(\phi,t)$}}

Now, we introduce another synchronization measure, namely, shifted phase distribution for the qubit system. The synchronization function $S(\phi,t)$, a quantitative measure that assesses the degree of phase locking inside the system \cite{PhysRevA.101.062104, PhysRevLett.121.053601}, is defined by integrating the $Q$ distribution function over the variable $\theta$ as follows:
\\

\begin{equation}
	S\left( {\phi ,t} \right) = \int_0^\pi  {d\theta \sin \theta Q\left( {\theta ,\phi ,t} \right)}  - {1 \over {2\pi }}= {{{\rho _{eg}}\left( t \right){e^{i\phi }} + {\rho _{ge}}\left( t \right){e^{ - i\phi }}} \over 8}
\end{equation}
\\

It is worth mentioning that $S(\phi,t)$ is zero if there is no phase synchronization, i.e., the uniform distribution of the $Q$-function. On the other hand, a nonzero value of $ S(\phi,t) $ implies the existence of phase locking. Specifically, a positive value of $S(\phi,t)$) indicates in-phase locking, while a negative value means anti-phase locking. The sign of $S(\phi,t)$ is expected to change as $\phi$ changes. However, the relationship between in-phase locking (anti-phase locking) and  $S(\phi,t)$ > 0 ( $S(\phi,t)$< 0) stays the same. This is because the difference between their phases determines whether the qubit and the modulation process are in phase or anti-phase. Phase locking transpires when the phase difference is almost zero, but anti-phase locking arises when the phase difference approaches $\pi$.\\
\\

\begin{figure}[h!]
	\centering
	\includegraphics[width=0.5\textwidth]{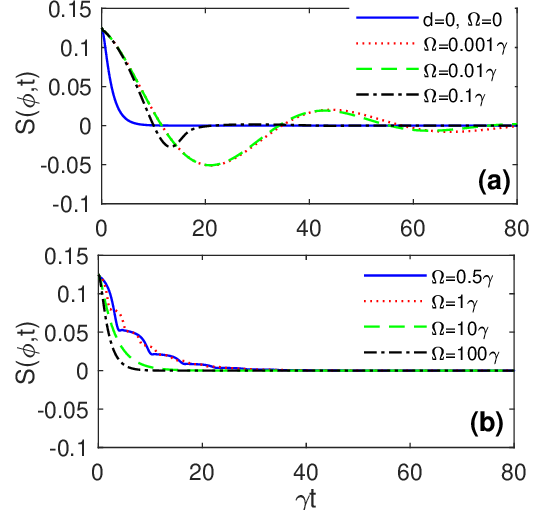}		
	\caption{The synchronization measure $S(\phi,t)$ versus scaled time $\gamma t$ for different values of the modulation frequency ($\Omega$). Solid-blue line in panel (a) corresponds to the situation in which frequency modulation is off, while for the other curves, the modulation amplitude is fixed at $d = 10\gamma$. other parameters are taken as $\phi$ = 0 and $\lambda= 3\gamma$ (the weak coupling regime).} \label{Fig5}
\end{figure}

\begin{figure}[h!]
	\centering
\includegraphics[width=0.8\textwidth]{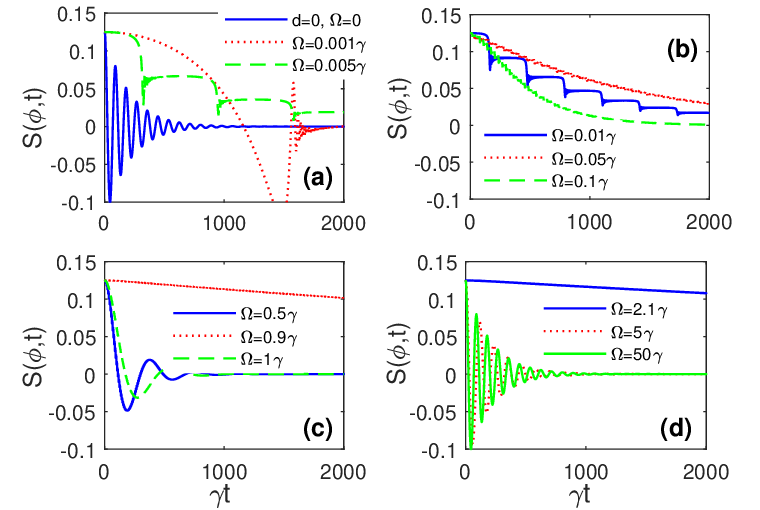}	
	\caption{The synchronization measure $S(\phi,t)$ versus scaled time $\gamma t$ for different values of the modulation frequency ($\Omega$). Solid-blue line in panel (a) corresponds to the situation in which frequency modulation is off, while for the other curves, the modulation amplitude is fixed at $d = 5\gamma$. other parameters are taken as $\phi$ = 0 and $\lambda= 0.01\gamma$ (the strong coupling regime).} \label{Fig6}
\end{figure}

Figure \ref{Fig5} helps us understand the synchronization dynamics shown in Figure \ref{Fig2}. Comparing these two figures can clarify the relationship between changes in the $Q$ distribution function and the value of $S$. We show the dynamics of $S(\phi,t)$ with respect to different values of modulation frequency. The first result is that, in fact, in both weak coupling and strong coupling regimes, no phase locking is expected in the absence of modulation frequency (i.e., the blue line in Figure \ref{Fig5}(a) and Figure \ref{Fig6}(a)). In Figure \ref{Fig5}(a), we can observe the rapid disappearance of the $S(\phi,t)$ function in the weak coupling regime. On the other hand, it can be observed that in the weak coupling regime, increasing the modulation frequency for both small and large values does not significantly affect on creating synchronization, and $S(\phi,t)$=0. This result suggests that in the weak coupling regime, the system has low sensitivity to changes in modulation frequency.
\\

Figure \ref{Fig6} shows the synchronization dynamics of $S(\phi,t)$ in the strong coupling regime. This regime is characterized by a more robust interaction between the two-level system (TLS) and its environment, leading to more complex behaviors. Without the modulation process, the $S(\phi,t)$ function initially exhibits damped oscillations induced by non-Markovian effects, specifically the backflow of information from the environment to the system. These oscillations eventually decay to zero, indicating a lack of synchronization in the long-term behavior of the system when no external modulation is applied.
However, the introduction of a modulation process significantly alters this behavior. As observed in Figure \ref{Fig6}(a), a minor modulation frequency induces phase locking ($S(\phi,t)$> 0) within the time interval $\gamma t$ < 300. This phenomenon demonstrates that even weak external modulation can temporarily synchronize the system's behavior. Nevertheless, for $\gamma t$ > 300, the phase locking deteriorates more rapidly, and $S(\phi,t)$ tends towards zero in the long term. This suggests that the initial synchronization induced by weak modulation is not sustainable over extended periods. The intriguing role of frequency dependence in controlling the system’s behavior is a key area for further exploration and understanding.
The behavior becomes particularly interesting for specific modulation frequencies. As noted earlier for the $Q$-function, and now explicitly shown in panels (c) and (d) of Figure \ref{Fig6} for $\Omega$ = $0.9\gamma$ and $\Omega$ = $2.1\gamma$ respectively, phase locking occurs and persists for substantially longer durations compared to other $\Omega$ values. This prolonged synchronization indicates the existence of resonant frequencies at which the system is particularly susceptible to external modulation. The stark contrast in behavior between different modulation frequencies highlights the system's sensitivity to the precise timing of external influences. This frequency dependence could be exploited in practical applications to control the system's behavior. The potential for controlling quantum systems through the careful selection of modulation frequencies is a significant practical implication of our research.
\\

\begin{figure}[h!]
	\centering
	\includegraphics[width=0.8\textwidth]{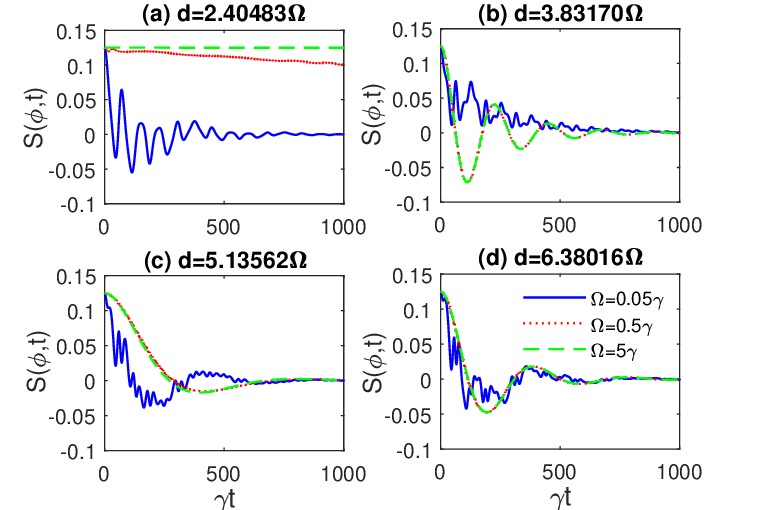}	
	\caption{The synchronization measure $S(\phi,t)$ versus scaled time $\gamma t$ for different values of the modulation frequency ($\Omega$). The panels correspond to various values of the modulation amplitudes: (a) $d = 2.40483\Omega$ [$J_0(2.40483) = 0$], (b) $d = 3.83170\Omega$ [$J_1(3.83170) = 0$], (c) $d = 5.13562\Omega$ [$J_2(5.13562) = 0$], (d) $d = 6.38016\Omega$ [$J_3(6.38016) = 0$]. Other parameters are taken as $\phi$ = 0 and $\lambda= 0.01\gamma$ (the strong coupling regime).} \label{Fig7}
\end{figure}

Figure \ref{Fig7} displays the dynamics of the synchronization $S(\phi,t)$ under strong coupling ($\lambda= 0.01\gamma$) when the ratio $d$/$\Omega$ is tuned such as to assume values that make the $n$-th Bessel function $J_n$ of Eq. (\ref{Eq5}) vanish. In Fig.\ref{Fig7}(a), we examine conditions where $d$/$\Omega$ is tuned to be the first zero of the Bessel function $J_0$ (the zeroth-order Bessel function) vanishes. In this case, for small values of modulation frequency, the synchronization function $S(\phi,t)$ initially shows a few damped oscillations. These oscillations are due to non-Markovian effects, interpreted as information returning from the environment to the system. After these initial oscillations, $S(\phi,t)$ tends towards zero, indicating the absence of stable synchronization ($S(\phi,t)$=0). However, as the modulation frequency increases, we observe a significant enhancement in phase locking, a key progress in our research. Yet, as shown in Fig.\ref{Fig7}(b-d), this behavior is not true for other considered values of the $d$/$\Omega$ ratio. These observations indicate that the relationship between modulation amplitude ($d$) and modulation frequency ($\Omega$) plays a crucial role in determining the synchronization in the system.
\\

\begin{figure}[h!]
	\centering
	\includegraphics[width=0.5\textwidth]{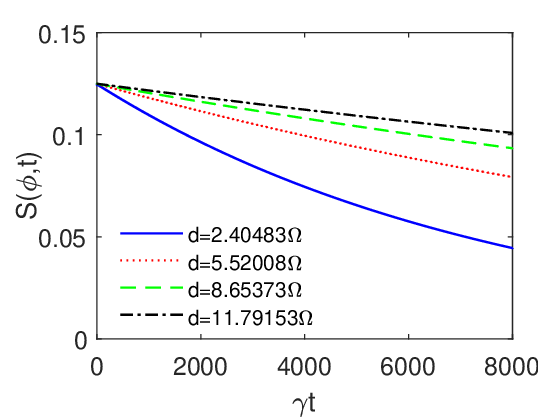}	
	\caption{The synchronization measure $S(\phi,t)$ versus scaled time $\gamma t$ for different values of the modulation amplitudes: (a) $d = 2.40483\Omega$ [$J_0(2.40483) = 0$], (b) $d = 5.52008\Omega$ [$J_0(5.52008) = 0$], (c) $d = 8.65373\Omega$ [$J_0(8.65373) = 0$], (d) $d = 11.7915\Omega$ [$J_0(11.7915) = 0$]. Other parameters are taken as $\phi$ = 0, $\Omega = 5\gamma$ (the modulation frequency) and $\lambda= 0.1\gamma$ (the strong coupling regime).} \label{Fig8}
\end{figure}

The results in Figure \ref{Fig7} motivate us to comprehensively examine setting $d$/$\Omega$ with other zeros of the zeroth-order Bessel function. This thorough analysis, presented in Figure \ref{Fig8}, allows us to draw and compare the dynamic behavior of the synchronization function for the first four zeros of the Bessel function of the zeroth order ($J_0$), highlighting the key differences and similarities. Here we consider $\lambda= 0.1\gamma$. By increasing the zeroth order of the zeroth-order Bessel function, it is seen that a better result for phase locking occurs, such that synchronization is lost later. Importantly, in all these cases, our calculations show that further increasing omega after $\Omega = 5\gamma$ gives the same result.

\section{CONCLUSION}
\label{sec.iv}

To summarize, this paper has considered the influence of frequency modulation on the quantum synchronization dynamics of a qubit coupled to the Lorentzian reservoir constituted by a leaky high-$Q$ cavity.  We have found that phase locking is negligible in the weak coupling regime, and the system has been shown minor sensitivity to changes in modulation parameters. However, the modulation process can lead to a significant phase-locking in the strong coupling regime. Our finding indicates that the precise selection of modulation frequency ($\Omega$) can meaningfully impact phase locking and hence synchronization dynamics. Moreover, as a key achievement, it has been demonstrated that the modulation amplitude-to-frequency ratio ($d$/$\Omega$) plays a very essential role in achieving a robust phase locking and, consequently, a durable synchronization in the system. So that when this ratio $d$/$\Omega$ was tuned to the zeros of the zeroth order Bessel function ($J_0$), synchronization could exist in the system for very long times. However, the result improves as the zero Bessel order becomes larger.
\\
This research deepens our understanding of synchronization dynamics in quantum systems under modulation and provides practical strategies for optimizing and controlling these systems. Our findings could have wide-ranging applications in quantum communications and quantum sensing devices. Importantly, this work sets the stage for future research, offering a promising direction for exploring the relationship between modulation parameters and synchronization behavior in more complex quantum systems.
\\

\bibliography{mybibb.bib}
\end{document}